 \documentclass[prx,twocolumn,secnumarabic,balancelastpage,longbibliography]{revtex4-1}

\usepackage[utf8]{inputenc}
\usepackage[T1]{fontenc}
\usepackage{bm,hyperref}
\usepackage{amsfonts} 
\usepackage{colortbl}
\usepackage{verbatim}

\usepackage{setspace}
\usepackage{lineno}
\usepackage{soul}

\usepackage{graphicx}   
\usepackage{import}                         
\usepackage{epstopdf}
\usepackage{amsmath} 
\usepackage{float} 
\usepackage{bm}
\usepackage{amssymb}
\usepackage{quotes}
\usepackage{indentfirst}
\usepackage{color}
\usepackage{transparent}
\usepackage{dcolumn}
\usepackage{braket}
\usepackage{multirow}
\usepackage{cancel} 
\usepackage{mdframed}
\usepackage{color}
\usepackage{bm}
\usepackage{dsfont}
\usepackage{slashed}
\usepackage{soul, color} 

\makeatletter
\newcommand*{\addFileDependency}[1]{
  \typeout{(#1)}
  \@addtofilelist{#1}
  \IfFileExists{#1}{}{\typeout{No file #1.}}
}
\makeatother

\begin{document}
\rmfamily
\title{A scalable platform for nanometer-scale quantum confinement}

\renewcommand{\figurename}{\textbf{Figure}}
\renewcommand{\thefigure}{\textbf{\arabic{figure}}}
\renewcommand{\hbar}
{\mathchar'26\mkern-9mu h} %
\renewcommand{\abstractname}{\vspace{-\baselineskip}}


\author{Christina~M.~Spaegele$^{1, \ddagger}$}
\email{spaegele.christina@gmail.com}
\author{Mehdi~Rezaee$^{2, \ddagger}$}
\author{Thomas~Werkmeister$^{2}$}
\author{Soon~Wei~Daniel~Lim$^{1,3}$}
\author{Kailyn~Vaillancourt$^{1}$}
\author{Joon-Suh~Park$^{1}$}
\author{Paul~Chevalier$^{1}$}
\author{Ido~Kaminer$^{4}$}
\author{Philip~Kim$^{2}$}
\author{Federico~Capasso$^{1}$}
\email{capasso@seas.harvard.edu}
\author{Michele~Tamagnone$^{5}$}


\affiliation{$^1$ Harvard John A. Paulson School of Engineering and Applied Sciences,\\ Harvard University, Cambridge, MA, USA}
\affiliation{$^2$ Physics Department, Harvard University, Cambridge, MA, USA}
\affiliation{$^3$ Department of Molecular and Cellular Physiology, Stanford University, Stanford, California 94305, USA}
\affiliation{$^4$ Solid State Institute and Faculty of Electrical and Computer Engineering, Technion – Israel Institute of Technology, Haifa, Israel}
\affiliation{$^5$ Fondazione Istituto Italiano di Tecnologia, Genova, Italy\\
$^{\ddagger}$ These authors contributed equally to the work.}


\begin{abstract}
\textbf{Abstract.} Overcoming the limitations of current nanofabrication techniques to achieve nanoscale feature sizes is essential for achieving new regimes of light-matter interactions at extreme frequencies and length scales. 
Here, we demonstrate a scalable nanofabrication platform capable of producing in-plane feature sizes down to 1.75~nm, pushing the boundaries of current top-down nanofabrication techniques. Using precise thickness control of atomic layer deposition (ALD) and employing widely spaced oxide nanofins, we transform conventional ALD into a surface structuring method that produces nanolaminates with sub-10~nm periodicities over large areas. 
The resulting nanostructures can be used as a one-dimensional gate array to control charge carriers in two-dimensional materials.
As an initial demonstration, we integrate the platform with graphene and perform electron transport measurements. In the presence of the gate array enabled by the nanolaminate, we observe satellite Dirac peaks consistent with band-structure modulation, suggestive of quantum-confinement effects.
Our platform paves the way for exploring previously inaccessible regimes of nanoscale light-matter interactions, holding significant promise for applications in short wavelength optics, electronics, and polaritonics.
\end{abstract}

\maketitle


Artificial materials with feature sizes commensurate with the wavelength of electromagnetic excitations have enabled a myriad of photonic and electronic devices, as well as the exploration of extreme regimes of light-matter interactions~\cite{Joannopoulos2008PhotonicCrystals,he2021moire,kadic20193d,yu2014flat}. Patterning a (quasi-)particle's environment at the wavelength scale leads to confinement-based quantization of electromagnetic modes and wavefunctions, thereby providing a means to control the particle's emission, absorption, or propagation~\cite{kittel2018introduction}. Periodic or quasi-periodic patterns generate potential landscapes that restrict the types and dynamics of electromagnetic modes allowed to exist within the nanostructured medium. For instance, for electrons in two-dimensional (2D) materials -- whose de Broglie wavelength is on the order of 1-100 nm -- quantum confinement effects start to become significant as the characteristic size of the system approaches these length scales ~\cite{kittel2018introduction}. 
Therefore, advances in fabrication techniques at and below the 10~nm scale are critically important for probing and harnessing effects in extreme ultraviolet nanophotonics~\cite{zhao2021recent}, free-electron physics~\cite{roques2023free}, valleytronics \cite{schaibley2016valleytronics} and next-generation 2D material-based devices~\cite{pham20222d}. 



\begin{figure*}
    \centering
    \includegraphics[scale=0.75]{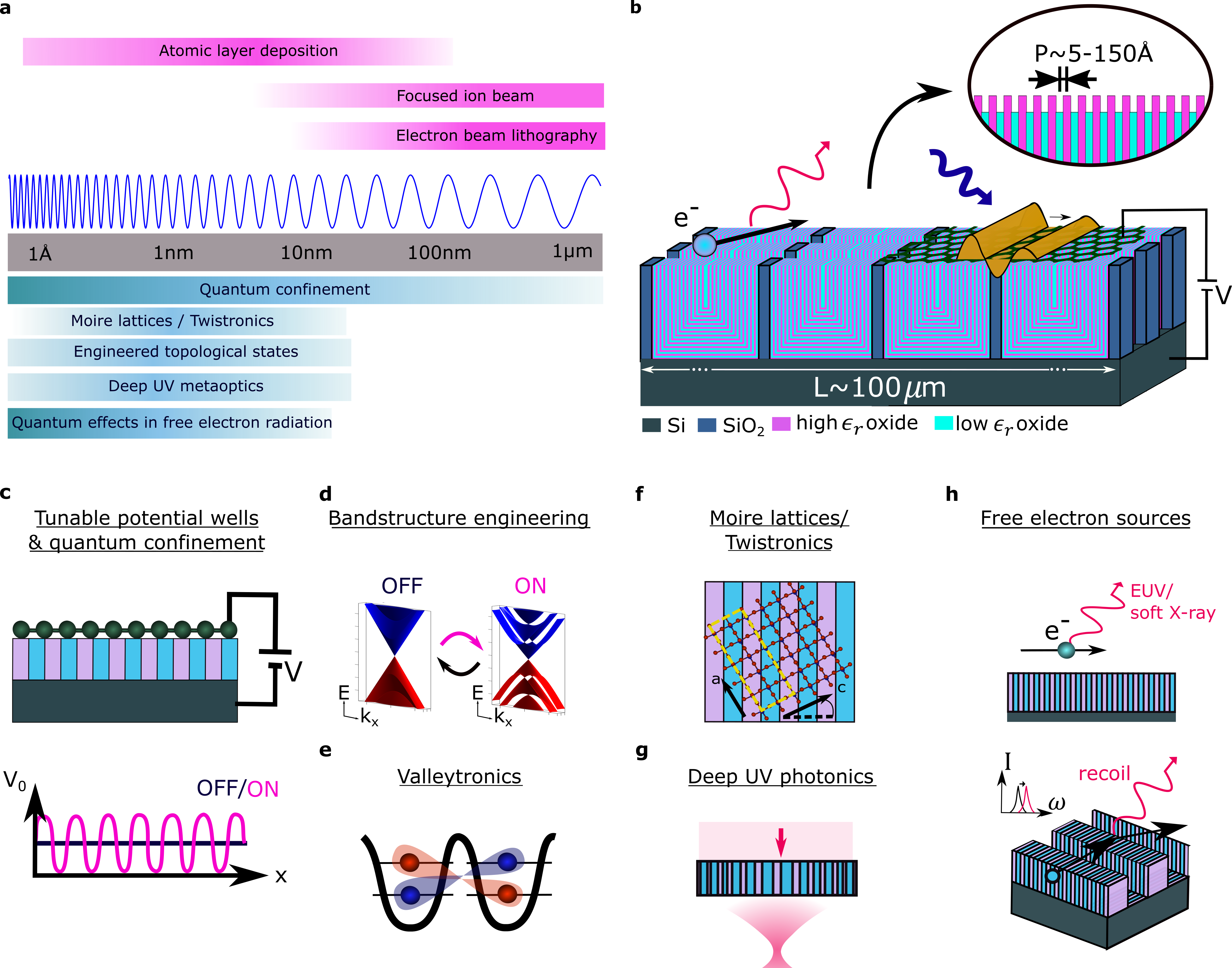}
    \caption{\textbf{Nanolaminate platform to attain extreme regimes of light-matter interactions.} (a) Comparison of the feature sizes accessible to various nanofabrication techniques (pink) against physical phenomena that become significant at corresponding length scales (blue). (b) Schematic of the proposed platform, relying on widely spaced oxide pillars to transform sub-nm thick layers grown by atomic layer deposition (ALD) into a nanolaminated surface.  (c–h) Applications of nanolaminates in extreme light-matter interactions: (c) When implemented with oxides of large permittivity contrast, applying a voltage between the platform and a 2D material creates tunable superlattices, able to tune the confinement of electrons in 2D materials. (d)  Band‐structure engineering: The imposed periodic potential modifies the 2D material’s electronic band structure. (e)  Artificially controlled quantum confinement at the sub‐10 nm scale facilitates quantum dot formation, valleytronics and exciton quantum confinement. (f) Because the patterning length scale is comparable to the moiré superlattice length scales observed in rotated 2D material stacks, the substrate could potentially serve as an additional degree of freedom for tuning electronic and optical properties in twistronics.  (g)  Photonic control in the deep ultraviolet (UV): an incident deep UV wavefront can be controlled by locally controlling nanoscale subwavelength features, of interest for photonic devices in hard to control wavelengths. (h) Free-electron radiation sources with quantum recoil: When a charged particle traverses above a periodic grating, it induces emission of radiation (e.g., Smith-Purcell effect). The proposed platform could be of interest to access unexplored regimes from soft X‐ray to near UV, reaching into the regions where interaction effects such as quantum recoil become pronounced.}
    \label{fig:concept}
\end{figure*}

Earlier foundational studies on surface quantum structures and superlattices extensively explored nanoscale periodicity for tuning material properties~\cite{kotthaus1991field}. More recently, active tuning of the band structure, transport, quantum confinement, and coupling of (quasi-)particles in 2D materials have been achieved through the introduction of a nanoscale periodicity in the system using etching~\cite{barcons2022engineering, sarkar2024sub}, thermal-probe scanning lithography~\cite{lassaline2021freeform}, applied strain~\cite{banerjee2020strain}, doping~\cite{sun2011towards}, patterning of the gating~\cite{huber2020gate} or insulation material~\cite{li2021anisotropic,forsythe2018band}, and twisted stacking of multiple 2D materials (``twistronics''~\cite{he2021moire}). Progress in nanoscale engineering has led to the emergence of novel quantum phenomena and enhanced material properties~\cite{park2009landau,brey2009emerging}, such as anisotropic electron transport~\cite{park2008anisotropic}, flat bands~\cite{shi2019gate,cao2018unconventional}, and the formation of replica Dirac cones~\cite{barbier2010extra}. Overall, however, efforts toward the nanoscale tuning of material properties remain limited by the absolute resolution of the current frontier in nanofabrication ~\cite{nano12162754,mrenca2023probing, pham20222d}.

Existing routes to sub-10~nm nanostructuring include optical lithography, electron- and ion-beam lithography, and atomically controlled growth methods such as molecular beam epitaxy and atomic layer deposition (ALD)~\cite{nano12162754,mack2008fundamental,vieu2000ebl,hoeflich2023fibroadmap,george2010ald,arthur2002mbe}. Yet this regime remains difficult to access in scalable, low-roughness photonic and electronic platforms. Top-down approaches are ultimately constrained by probe resolution and roughness at the smallest dimensions; electron- and ion-beam methods face additional limits from resist resolution, proximity effects, and intrinsically slow serial writing over large areas~\cite{mack2010ler,okazaki2015opticalebl,hoeflich2023fibroadmap}. By contrast, ultra-thin layer growth methods offer atomic-scale thickness control but generally do not enable direct surface patterning~\cite{george2010ald,johnson2014aldbrief,arthur2002mbe}. Consequently, scalable material platforms that combine sub-10~nm dimensions with in-plane patterning remain scarce, leaving many photonic and electronic regimes largely unexplored.

Here, we demonstrate a new nanofabrication technique to overcome the aforementioned challenges, enabling on-chip surface feature sizes (half the in-plane periodicity) down to the single-nanometer scale. 
Our technique leverages the precise layer thickness control offered by ALD, a thin film coating technique that allows for the deposition of materials one atomic layer at a time, to produce in-plane nanometer and sub-nanometer features that are wafer scalable (Figure~\ref{fig:concept}(b)). By utilizing widely spaced oxide nanofins, our platform repurposes ALD into a surface structuring method capable of producing nanolaminates with feature sizes experimentally demonstrated  down to 1.75$\mathrm{nm}$.

We extensively characterize the nanolaminates' topography, feature sizes, and material contrast using atomic force microscopy (AFM), scanning- and transmission electron microscopy (SEM and TEM, respectively), and energy dispersive X-ray spectroscopy (EDS). 
Our method has the potential to push quantum confinement to hitherto unachievable energy scales.
We demonstrate the potential of our platform for band structure engineering of 2D material systems by placing graphene across the patterned surface of ~6.25~nm feature size nanolaminates and characterizing their gating effect on electron transport. 
We observe the emergence of satellite Dirac peaks consistent with theoretical expectations, offering preliminary evidence for quantum-confinement behavior at these scales.
We anticipate that our platform will enable the exploration of novel regimes of light-matter interactions at the sub-10~nm length scale with potential applications in high-frequency optics, electronics, and polaritonics.

\section*{Motivation and prospects for the nanolaminate process}

We first outline how the proposed nanolaminate fabrication technique could serve as a platform for exploring extreme regimes of light-matter interactions, across nanophotonics, 2D material physics, and electron microscopy. 
Our method is compatible with feature sizes below 10~nm, a regime in which the quantum confinement effects highlighted in Figure~\ref{fig:concept}(a) become prominent. This suggests a possible route toward accessing unconventional and typically hard-to-reach regimes of light-matter interactions, which are of interest both from a fundamental physics standpoint and for electromagnetic device applications.

Due to the wide range of dielectric materials (and hence permittivities) that can be deposited with ALD, the proposed nanolaminate platform appears well-suited for active tuning of 2D material band structures. When integrated with (semi-) metallic 2D materials, such as graphene, and a conductive substrate, the system can be treated as a capacitor comprising alternating dielectric regions across the two plates. A voltage $V$ applied between the substrate and the grounded 2D material induces a voltage-tunable, periodic potential onto charges in the 2D material (Figure~\ref{fig:concept}(c)). The depth of the potential increases with the relative permittivity difference of neighboring materials. The induced structured lattice of potential wells or barriers (which we refer to as ``superlattice''~\cite{sun2011towards}) could give rise to phenomena such as anisotropic band flattening, the formation of additional satellite Dirac cones, and fractal Hofstadter spectra under magnetic fields \cite{hofstadter1976energy}. These superlattice effects may enable control over electron transport and carrier dynamics ~\cite{forsythe2018band, li2021anisotropic}, including in regions of the Brillouin zone and energy bands that are otherwise difficult to access. The applied voltage can act as a tuning knob for the topography of the band structure (Figure~\ref{fig:concept}(d)), thus controlling the optoelectronic properties of 2D materials.\

Furthermore, the scalability of this method to ultra-small feature sizes may enable new avenues in excitonic quantum confinement, particularly relevant for transition-metal dichalcogenides, where potential widths <10~nm are required to achieve confinement \cite{thureja2022electrically}. By exploiting the dielectric-defined potentials and 2D prepatterning, the nanolaminated substrate  may host deterministically positioned, voltage-tunable (inter- and intralayer) excitonic quantum dots. This capability may open opportunities in quantum photonics, including single quantum emitters with well-defined spatial arrangements. Additionally, spatially controlling the excitonic potential energy and density of inter- and intralayer excitons at such otherwise hard-to-access length scales could support exotic quasiparticle phenomena such as topological excitons and exciton-polaritons, facilitating novel quantum many-body phases~\cite{perczel2017topological, PhysRevLett.118.147401, ciarrocchi2022excitonic}(Figure~\ref{fig:concept}(e)). 

Because the patterning length scale is comparable to the moiré superlattice length scales observed in rotated 2D material stacks (10~nm)~\cite{cao2018unconventional}, the nanolaminated substrate can serve as an additional degree of freedom for tuning electronic and optical properties in twistronics. For even smaller length scales approaching sub-nanometer periodicities or in 2D materials exhibiting directional anisotropy (e.g., MoO$_3$~\cite{kim2023plane}), this approach could potentially allow for rotation angle dependent system responses, analogous to methods used in twistronics (Figure~\ref{fig:concept}(f)). Such patterned substrates can potentially enhance valley-dependent phenomena and control valley polarization, leveraging substrate-induced symmetry breaking or modulation effects that are pivotal in valleytronics~\cite{ciarrocchi2022excitonic, yu2017moire,schaibley2016valleytronics}.

The $<10$~nm length scale also underpins effects in nanophotonics and electron-light interactions. The realization of patterned surfaces with feature sizes at these length scales may facilitate the development of deep ultraviolet metaoptics~\cite{ossiander2023extreme, zhao2021recent} with important applications in extreme UV lithography at 13.5~nm~\cite{wu2007extreme} (Figure~\ref{fig:concept}(g)).  At these extreme length scales, observation of quantum recoil~\cite{tsesses2017light, huang2023quantum} from free electrons is also facilitated~\cite{tsesses2017light} (Figure~\ref{fig:concept}(h)). The exploration of these regimes of extreme light-matter interaction motivates the development of nanofabrication methods with sub-10~nm feature sizes. 
\begin{figure*}[!ht]
    \includegraphics[scale=0.9]{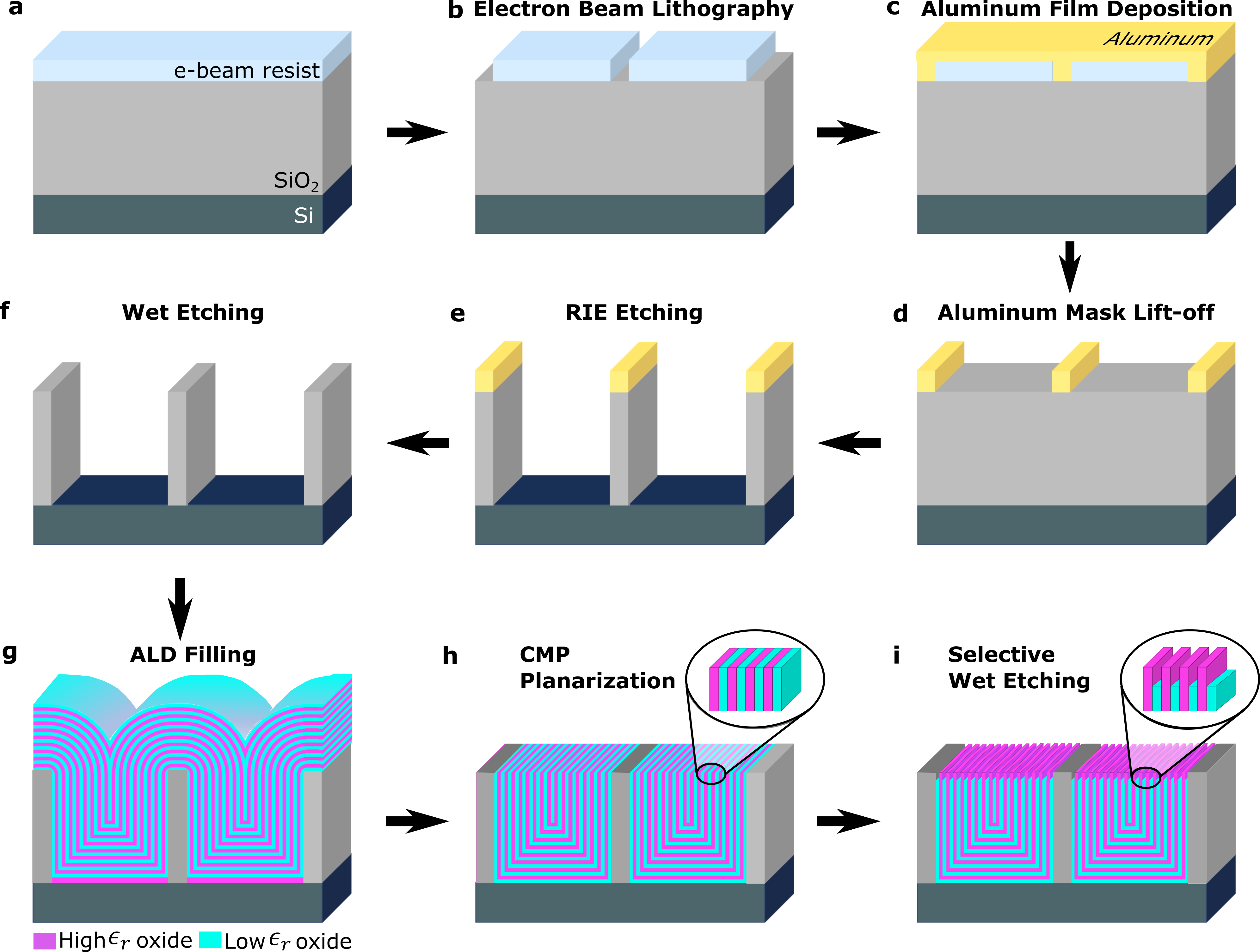}
    \caption{\textbf{Fabrication of the nanolaminate structure.} (a-f) Creation of the trenches between oxide nanofins. An aluminum (Al) etch mask is patterned on a thermal oxide silicon wafer using (a-b) electron beam lithography, (c) Al thin-film deposition using an electron beam evaporation,  and  (d) subsequent lift-off process. (e) The oxide nanofins are formed using a reactive ion-etch, and (f) thinned out using a wet etch process which simultaneously removes the Al etch mask. (g) The trenches are then filled with alternating layers of two different materials (hafnium dioxide (hafnia) and aluminum oxide), using ALD. (h) The ALD-filled substrate is then planarized using chemical mechanical polishing (CMP), exposing the alternating material film in parallel to the trench orientation. This step ensures the 2D material can be placed in close contact with the nanopatterned substrate. (i) Finally, the platform's material contrast is amplified by selectively etching away one of the materials. In this case, alumina is partially etched using a buffered oxide etch (BOE), leaving behind raised ridges of hafnia.}
    \label{fig:fab}
\end{figure*}

\begin{figure*}[!ht]
    \includegraphics[scale=0.79]{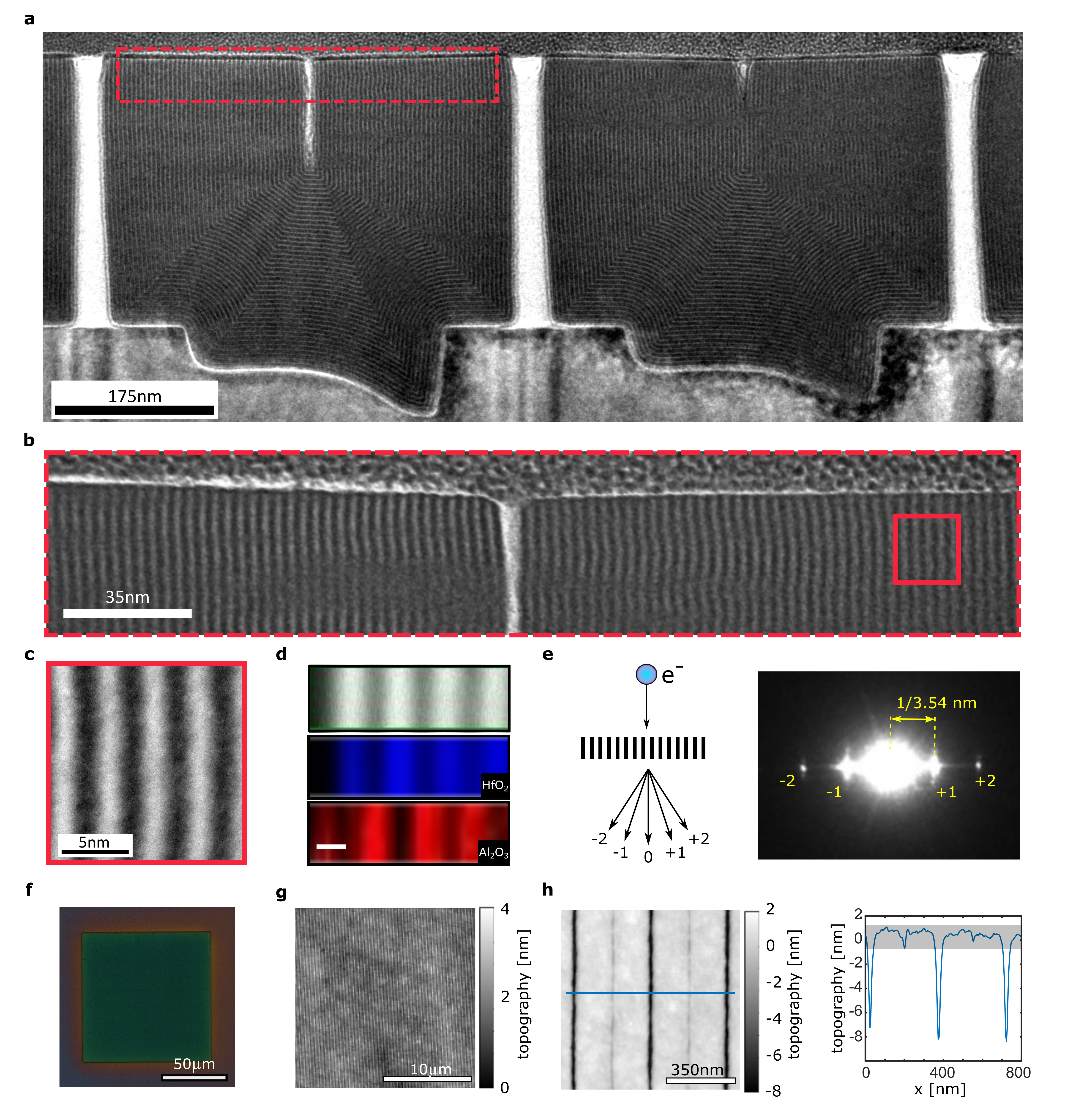}
    \caption[Characterization of nanolaminate superlattices.]{\textbf{Characterization of nanolaminate superlattices}. We characterize a square region of side length 100 $\mathrm{\mu}$m with a feature size of 1.75~nm (periodicity 3.5~nm). (a) TEM image of two neighboring super-periods.  (b) Zoomed-in TEM images of the lamella of dashed area in (a). (c) Zoomed-in TEM image of red boxed area in (b), showing well separated material layers of periodicity $P=3.5$~nm. (d) Material contrast is further highlighted using TEM energy dispersive X-ray spectroscopy (EDS). The elemental maps below the STEM image show the distributions of Hf (from $\mathrm{HfO_2}$, blue) and Al (from $\mathrm{Al_2O_3}$, red). The clear periodic variation in elemental intensity confirms the controlled multilayer composition of the superlattice structure. Scale bar: 1.75~nm. (e) Electron diffraction patterns from the superlattice, obtained using a 200~keV electron beam, display clear diffraction orders. The spacing between the electron diffraction pattern  validate the periodicity predicted by the STEM image up to their measurement uncertainty with $P = 3.54 \pm 0.01$~nm. (f) Optical image of the full 100$\times$100~$\mu$m scale area. (g-h) AFM surface topography measurement and cross-sectional cut, showing a flatness of 4~nm over 20 $\mu$m, with a sub-2~nm height variation in between adjacent oxide nanofins.}
    \label{fig:ebeam}
\end{figure*}

\vspace{-2mm}

\section*{Fabrication process and characterization of sub-10~nm nanolaminate structures}

We now elaborate on the fabrication of nanolaminates with nanometer-scale feature sizes, reaching down to 3.5~nm periodicity. Figure~\ref{fig:fab} illustrates the fabrication process of the proposed platform. We can realize nanometer-scale in-plane resolution by ``redirecting'' the defining ALD feature of atomically-flat layers with controlled nanoscale thicknesses. Instead of depositing ALD layers onto a flat surface, we first introduce widely spaced nanofins onto a substrate, then perform ALD on the structured surface. These nanofins are etched into a thermal oxide layer of a silicon wafer (thickness $d=285$~nm) using an aluminum etch mask. The mask is created using standard lift-off techniques, defining the positions of the oxide nanofins on the substrate through electron beam lithography (Figure~\ref{fig:fab}(a-b)), followed by electron beam evaporation of an aluminum layer (Figure~\ref{fig:fab}(c)) and subsequent liftoff in a solvent stripper (Figure~\ref{fig:fab}(d)). Due to the wide spacing and low spatial resolution requirements of the nanofins, this step is suitable for large-area electron beam fabrication (e.g., at the cm scale). While in this setting the nanofins are chosen to be one-dimensional lines spaced by a pitch length $P=350$~nm, they can have free-form shapes for 2D surface control beyond parallel grating lines. The nanofins are then defined using reactive-ion etching (Figure~\ref{fig:fab}(e-f)).

The gaps so created are subsequently filled up conformally with alternating layers of ALD (Figure~\ref{fig:fab}g). Hafnium dioxide (HfO$_2$) and aluminum oxide (Al$_2$O$_3$) were chosen for their pronounced difference in relative permittivity ($\epsilon_\text{HfO$_2$} = 25$ and $\epsilon_\text{Al$_2$O$_3$} = 9$). 
The material that protrudes above the oxide teeth is then removed using chemical mechanical polishing (CMP) using a soft material cloth and a solution comprising 20~nm-small SiO$_2$ colloids (Figure~\ref{fig:fab}(h)). This planarization step is crucial to ensure good contact with the 2D materials (i.e., no ``bridging'' between super-periods -- see supplementary information (SI) for a detailed description of this step). Lastly, the material contrast can be further enhanced through a selective wet etch (Figure~\ref{fig:fab}(i)), removing the low permittivity material (in case of HfO$_2$ and Al$_2$O$_3$ a buffered oxide etch is suitable). The conductive silicon substrate supporting these structures can serve as the back gate once 2D materials are deposited on the top surface. For optical characterization applications, alternative substrates such as undoped silicon can be used to minimize optical losses. Alternatively, the Si substrate can be completely removed using the Bosch etch process, providing flexibility depending on specific application requirements.

While periodically spaced oxide nanofins also have a characteristic super-period corresponding to the pitch length $P=350$~nm, this super-period is one to two orders of magnitude larger than the small-scale nanolaminate periods, making the effect of the large super-period negligible in the short wavelength regions of interest. The distance between nanofins can be further increased by using thicker thermal oxide layers and thus deeper trenches, further decreasing the impact of the oxide nanofins if necessary.

\subsection*{Revealing sub-nanometer periodicities with electron diffraction}
 
Figure~\ref{fig:ebeam} shows an example of a fabricated device of area 100 $\mu$m $\times$ 100 $\mu$m  with feature sizes of 1.75~nm (Figure~\ref{fig:ebeam}(e)). As such features are too small to resolve with conventional scanning electron microscopy, a lamella was cut out of the grating to image the periodicity using TEM (see Figure~\ref{fig:ebeam}(a-c)), revealing high material contrast using TEM energy dispersive X‐ray spectroscopy (EDS) (Figure~\ref{fig:ebeam}(d)).

To further characterize our platform and perform initial tests for its applications in electron beam manipulation, we measure electron diffraction in a TEM. When a collimated 200~kV electron beam ($\lambda_{\mathrm{deBroglie}\ }= 2.5 \ \mathrm{pm}$) is impinging at normal incidence on the superlattice with period ${P}\approx 3.5\ \mathrm{nm})$, the superlattice behaves as a diffraction grating, diffracting the n$^{th}$ diffraction order by an angle $\alpha_n=0.72n ~\mathrm{mrad}$ with $n$\ $\in[\pm1,\pm2,\ldots ]$. We observe two clear diffraction orders (Figure~\ref{fig:ebeam}(e)) which are evidence for the long-range order of the structure, complementing the short-range order evident in the STEM and a promising indicator for future experiments involving fast electrons interacting with the superlattice~\cite{roques2023free}. The peak-to-peak distances $(P_{\pm1}=0.565\pm0.003~{\mathrm{nm}}^{-1}$ and  $P_{\pm2}=1.128\pm0.002~{\mathrm{nm}}^{-1}$, respectively), confirm the periodicity to be $P=3.54\pm0.01$~nm (see SI for details). The surface flatness was measured using atomic force microscopy (AFM), showing a flatness of 4 nm over 20 $\mu$m, with a sub-2 nm height variation in between adjacent oxide nanofins (Figure~\ref{fig:ebeam}(f-h)). 
	
 As the maximum potential depth achievable is limited by the breakdown electric field of the oxides used, a breakdown test was performed by applying a voltage between the doped back gate and a metal patch deposited on top of the fabricated surface (see SI). The breakdown electric field was determined to be 0.35~V/nm, corresponding to a breakdown voltage of 100~V. 

\begin{figure*}
    \centering
    \includegraphics[scale=0.7]{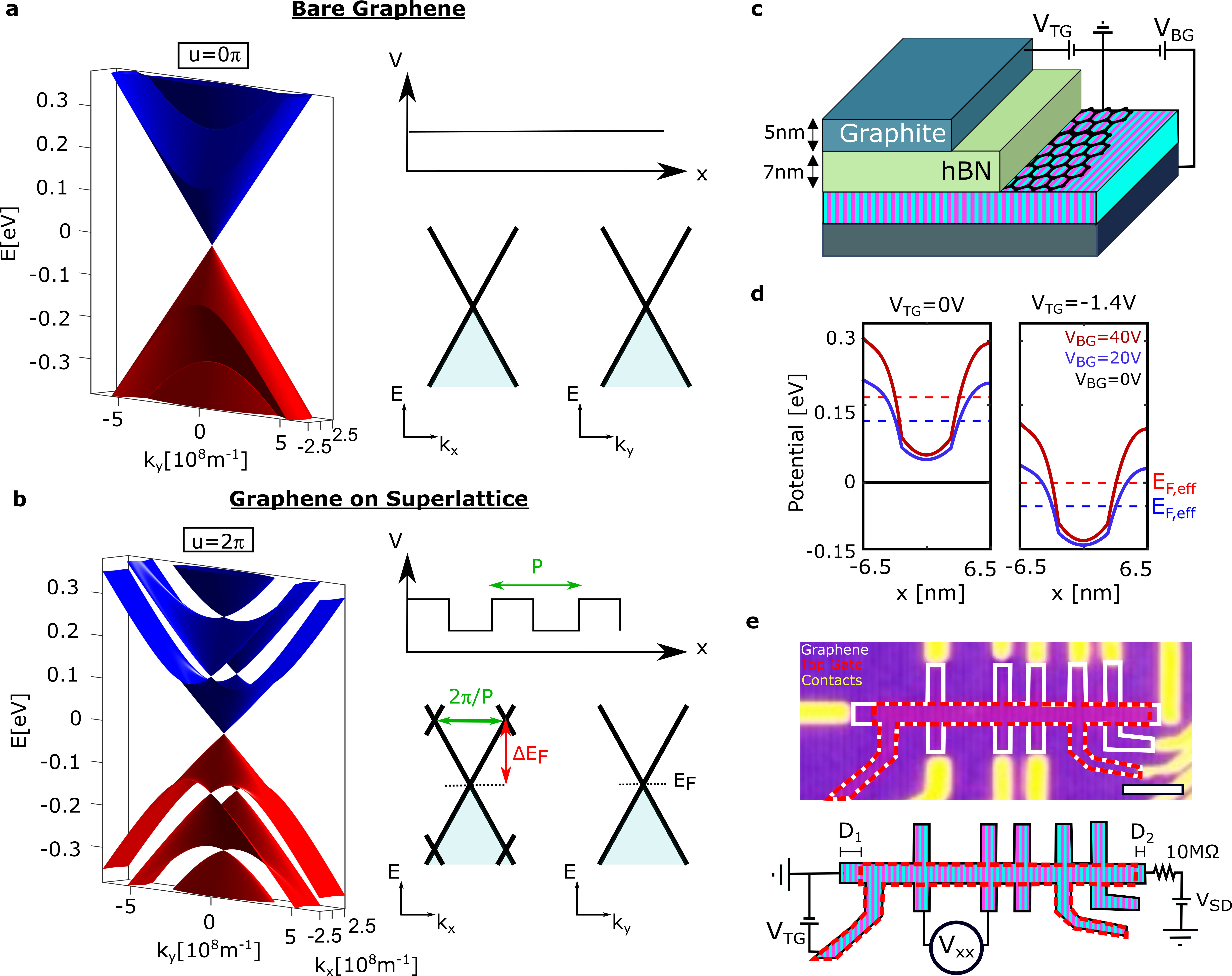}
    \caption{\textbf{Quantum confinement in graphene with nanolaminate superlattices.} (a) The band structure of bare graphene is shown in a mini-Brillouin zone (MBZ) with dimensions \(\pi/L_x\) and \(\pi/L_y\), where \(L_x = 12.5~\text{nm}\) and \(L_y\) is the atomic spacing. In the absence of a spatially varying potential (\(V(x) = \text{const.}\), top right), the band structure exhibits the characteristic cone-like dispersion near the Fermi level. This is further illustrated in the bottom right schematic, which shows unfolded cross-sections through the MBZ, emphasizing the linear Dirac dispersion in both \(k_x\) and \(k_y\) directions.
 (b) The band structure of graphene can be changed by applying a periodic potential $V(x)$ of period $P$. With increasing potential depth, the band structures develops satellite Dirac cones at the edge of the plotted mini‐Brillouin zone, corresponding to $k_x=\pi/P$. These features can be connected to a characteristic Fermi level distance $\Delta E_F$ and used as a characteristic sign of successful band structure engineering. (c) Schematic of the device used to characterize the band structure change displayed in (a). A stack consisting of graphene, hexagonal boron nitride (hBN) and graphite is placed on top of the laminate platform of feature sizes of 6.25~$\mathrm{nm}$. The potential depth of the superlattice can be tuned by an applied voltage with respect to the doped silicon substrate. hBN serves as the insulating dielectric layer of a top graphene gate, which allows to tune the Fermi level of graphene using an applied top gate voltage $V_\text{TG}$. (d) Simulated effect of the applied voltages on the potential well depth. The potential well depth $V_\text{0}$ is increased with the applied back gate voltage $V_\text{BG}$. This increase is accompanied by an overall offset that can be compensated by the application of a top gate voltage $V_\text{TG}$. (e) Optical image and schematic of the used Hall bar design. The resistance of the device is measured by applying a constant current $I_\text{SD}$ and measuring the voltage $V_{xx}$ between the side contacts of the Hall bar. Scale bar 3~$\mu$m.}
    \label{fig:device}
\end{figure*}

\section*{Band structure engineering and quantum confinement of 2D materials}

We now present a proof-of-concept demonstration of the potential of our platform in controlling the band structure of 2D material-based devices. Specifically, we utilized the superlattice to control the band structure of graphene-based systems by using an applied back gate voltage to tune the strength of the superlattice band structure modulation. 

The response of graphene to the superlattice in the device shown in Figure~\ref{fig:concept}(c) can be modeled as a one-dimensional (1D) potential well of depth $V_{\text{max}}$ altering the site-dependent eigenenergies of electrons. The corresponding band structure of graphene can be determined as the eigenvalues of the Hamiltonian $\mathcal{H}$ using a standard $k\cdot p$ method~\cite{li2021anisotropic} with 
$\mathcal{H}=v_F~\vec{\sigma}\cdot\vec{k}+I_2V_\text{1D}\left(x\right)$,
where${\ v}_F$ is the Fermi velocity of unpatterned graphene, $\sigma$ is the vector of Pauli matrices $(\sigma_x, \sigma_y)$, $\vec{k} = (k_x, k_y)$ is the wave vector measured from the Dirac point in the graphene Brillouin zone, $I_2$ is the 2$\times$2 identity matrix, and $V_{1D}\left(x\right)\ $is a Kronig-Penney-type potential of potential depth $V_\text{max}$. The resulting band structure can be tuned by changing the potential strength.

Figure~\ref{fig:device}(a) shows the effect of a potential well of periodicity $P$ on the band structure of the graphene superlattice system. Graphene itself exhibits a characteristic Dirac cone symmetric around $E_F=0$. When a square potential of period $P$ is applied, the band structure changes, with characteristic satellite Dirac points (SDPs) appearing at $k_x=\pi/P$ for a potential depth of $V_0=2\pi \hbar v_F/P$~\cite{li2021anisotropic} (Figure~\ref{fig:device}(b)). For decreasing periodicity, the required potential depth $V_0$ for the SDPs to appear increases, which can be understood as a consequence of the SDPs moving to higher energies for smaller periodicities. This relation is characterized through the dimensionless number $u=V_0 P /(\hbar v_F)$, with $u=2\pi$ for the first order SDP. These SDPs are created since there is a gap opened at the Brillouin zone edge for all $\vec{k} $ vectors except for $\vec{k} =\pm \pi/P ~  \hat{x}$. The SDP arises at $\vec{k} =\pm \pi/P ~  \hat{x}$ is topologically protected by the chirality of the Dirac cone band structure (detailed discussion on the topological protection can be found in the SI, including Figure~S5 for graphene band structures for different potential depths). More generally, higher-order SDPs appear at parameter points satisfying $u = 2\pi m$ with $m\in \mathbb{Z}$. In our experiment, these characteristic satellite features serve as a signature of successful quantum confinement and band structure engineering through the proposed platform.

\begin{figure*}
    \includegraphics[scale=0.7]{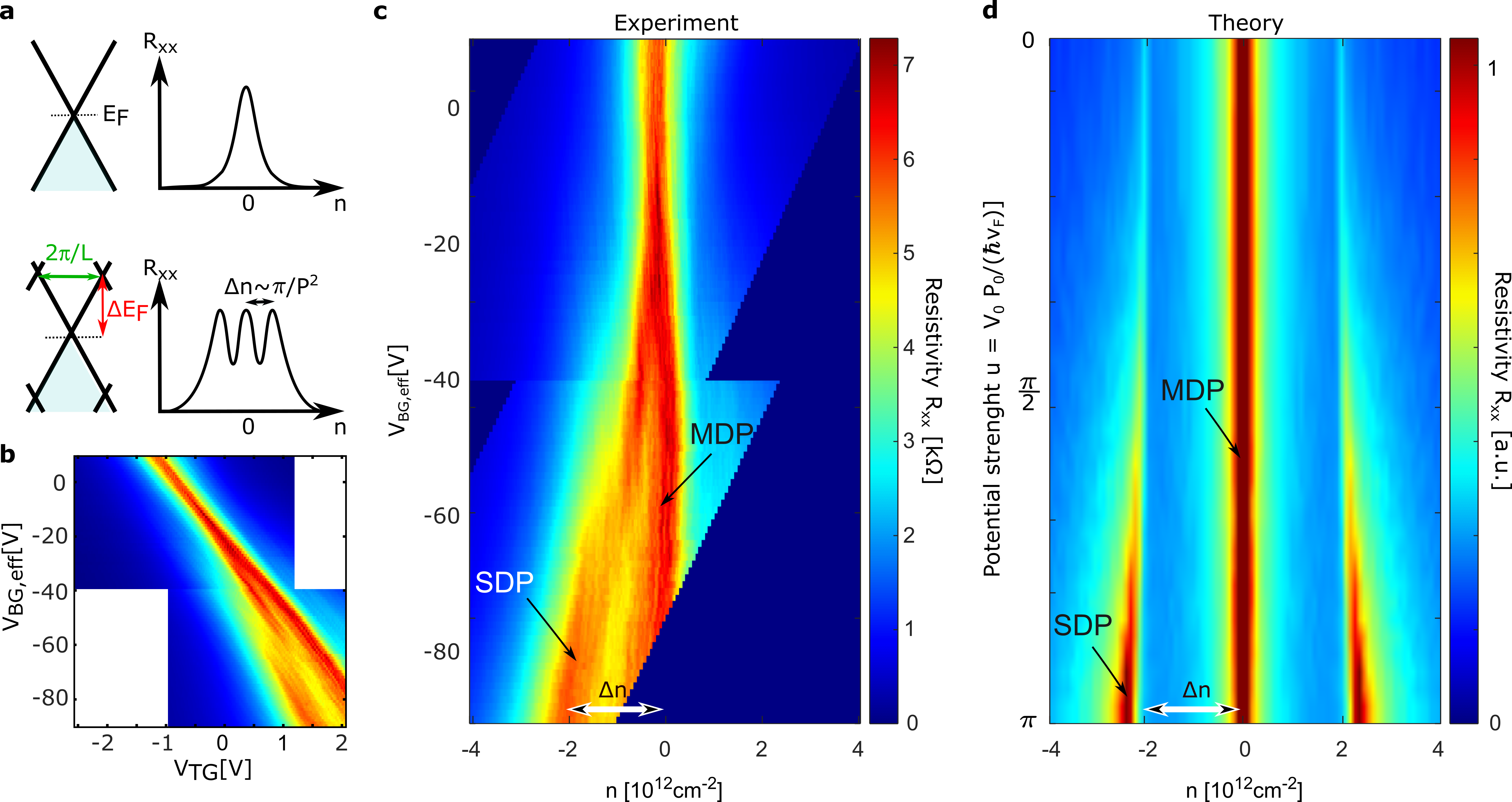}
    \caption{\textbf{Signatures of quantum confinement in graphene with nanolaminate superlattices.} (a) When the Fermi level (or equivalently the carrier density) is swept across the Dirac cone through an increase of $V_\text{TG}$, the measured Hall bar resistance peaks around $n=0$ due to the reduced carrier density at the main Dirac point (MDP). Similarly, the resistance of graphene becomes maximal around the satellite Dirac points (SDP) of the band structure-engineered graphene superlattice system. This leads to the appearance of additional resistance peaks at distance $\Delta n \propto \pi/P^2$. (b) Resistance measurements as a function of top-gate and effective back-gate voltages. As the back gate voltage $V_\text{BG,eff}$ is decreased, the resistance peak drifts towards higher top-gate voltages, due to the Fermi level shift $\Delta E_F$ described in Figure~\ref{fig:device}(d). For large negative $V_\text{BG,eff}$ $\sim-40~$V, additional side peaks corresponding to satellite Dirac cones emerge due to the increase in the potential well depth (proportional  to the back gate voltage) seen by the electrons. (c) Zoom into the resistance measurement from (b) as a function of estimated carrier densities (see SI for conversion method from voltage to carrier density). The difference in charge carrier density between the peaks is ($\Delta n \approx 1.8 \times 10^{12}$~cm$^{-2}$ corresponding to a superlattice with feature sizes of approximately 6.25~$\mathrm{nm}$ ($P=12.5~\mathrm{nm}$). (d) Simulated resistance of the graphene superlattice system for a feature size of 6.25~nm, showing the appearance of characteristic resistance peaks. The observed $\Delta n$ matches the experimental data.}
    \label{fig:signature}
\end{figure*}
\subsection*{Characterizing the band structure using electron transport measurements}
 SDPs can be revealed with an electron transport measurement of graphene's doping-dependent resistance. As the density of states is minimal at SDPs, the resistance of graphene will peak when the Fermi level passes them, making their experimental observation a robust indicator of band structure engineering~\cite{li2021anisotropic,mrenca2023probing}. Figure~\ref{fig:device}(c) shows the 2D material stack used in this work. It consists of the laminate-patterned back gate and a hBN/graphite based top gate (7~nm and 5~nm thickness, respectively). For this specific design we have chosen a nanolaminate feature size of 6.25~nm (periodicity $P=12.5$~nm). 
The back gate voltage $V_\text{BG}$ can control the superlattice potential depth and hence tune the band structure of the graphene superlattice system (Figure~\ref{fig:device}(d)). The graphite top gate is utilized to tune the Fermi level of graphene through an unpatterned hBN top dielectric layer, enabling probing of the satellite Dirac cones by sweeping the Fermi level through the satellite features. The top gate is additionally able to compensate for the Fermi level offset induced by the back gate.
 The resistance of the device is measured using a Hall bar, i.e., a narrow graphene strip (length $W=15~\mu \mathrm{m}$, width $w=1.1~\mu \mathrm{m}$) surrounded by contacts to measure the resistance of the device. By applying a constant source-drain current $I_\text{SD}$ and measuring the longitudinal voltage $V_{xx}$ (see Figure~\ref{fig:device}(e)) between the side contacts of the Hall bar, it is possible to measure the resistance of the device and track its changes for variations of the top gate voltages (i.e., the Fermi level). 

Figure~\ref{fig:signature} shows the experimental and simulated resistance of the performed transport measurement. 
 We observe features consistent with the presence of SDPs with our measurement in Figure~\ref{fig:signature}(b), showing the measured resistance of graphene, dependent on the back- and top gate voltages.  The diagonal shift of the main peak originates in the Fermi level offset described in Figure~\ref{fig:device}(d). In our experiment, the SDP only appears on one side of the main Dirac cone, an asymmetry that is common for electron transport-based superlattice characterizations using Hall bars~\cite{mrenca2023probing}. 
 This can be explained by the formation of \(p\)-\(n\) junctions between the contacts and the dual-gated region of interest,  such as at the single-gated regions marked in Figure~\ref{fig:device}(e) by the distances \(D_1\) and \(D_2\) between the gold contacts and the top gate. These $p$-$n$ junctions form when the single-gated regions are relatively hole-doped compared to the dual-gated region, reducing contact transparency and leading to incomplete equilibration and degradation of the measured voltage signals~\cite{delagrange2024residual}.



To verify that the observed discontinuity of the resistance peak arises from the nanolaminate periodicity, we convert the top-gate voltage to carrier density $n$ (Figure~\ref{fig:signature}(c); see SI for details). Both the effective back-gate voltage $V_{\text{BG,eff}}$ and $n$ are referenced to offsets $V_{\text{BG,0}}$ and $n_0$, chosen by imposing symmetry of the zero-field transport about $V_{\text{BG,eff}}=0$ and $n=0$. Using the measured average lattice period of the nanolaminate ($P=12.65\,\text{nm}$; see cross-section in the SI), the expected density spacing is $\Delta n \approx \pi/P^2 \approx 1.96\times 10^{12}\,\text{cm}^{-2}$, in good agreement with both the data and our theory (arrows in Figure~\ref{fig:signature}(d)). We believe the initial splitting at $V_{\text{BG,eff}}\approx -50$V likely originates from the larger 350~nm oxide superperiod, while the subsequent distinct satellite feature reflects the intended 12.5~nm laminate periodicity (see SI, Figure~S10). 

To summarize, the emergence of satellite Dirac cones revealed by resistance measurements, with spacings corresponding to a charge density modulation and confirmed by our theory, all indicate that our nanolaminate can successfully engineer the band structure of 2D materials placed on the surface.


\section*{Discussion and outlook} 

In summary, we introduced a nanofabrication platform capable of implementing in-plane feature sizes well below 10~nm, while remaining scalable to large areas (demonstrated 100~$\mu$m, and in principle wafer-scalable). We have observed satellite peak signatures in electron transport measurements of graphene placed on the nanolaminate. These constitute preliminary evidence for the potential of our method to achieve quantum confinement at the few-nanometer scale. While the transport measurements focused on superlattices with 6.25 nm feature sizes, we have also realized samples with much smaller feature sizes (1.75 nm, see Figure~\ref{fig:ebeam}). 

To the best of our knowledge, the nanofabrication technique presented here achieves the smallest feature sizes reported so far in the literature for fabrication techniques that can be scaled to hundreds of $\mu$m and more. Conventional device fabrication methods either (1) fail to access sub 10~nm scale, due to the minimum feature sizes accessible to conventional nanofabrication techniques; (2) lack tunability since they are constrained by the atomic arrangement and potential depth of the stacked two-dimensional materials; or (3) are not scalable to larger areas. These limitations also often constrain accessible regimes of engineering light-matter interactions ~\cite{mrenca2023probing, pham20222d}. For instance, while cleaved edge fabrications~\cite{pfeiffer1993cleaved} can reach similar feature sizes, they are limited to smaller areas and do not allow for back gate gating, and hence, do not permit band structure tuning once fabricated. Furthermore, creating a grating of width 100~$\mu$m using the cleaved edge method would require 500 times more material to be deposited than the proposed method (which requires 175-200~nm and is independent of the area). Other techniques are either constrained to larger feature sizes (top-down patterning) or are restricted to hexagonal lattice symmetries and smaller potential depth with limited scalability (Moir\'e lattices in twisted 2D materials~\cite{forsythe2018band}). 

While the proposed platform employs parallel grating nanofins, these nanofins are not confined to one-dimensional structures and can adopt any (closed or open) shape, provided that neighboring oxide boundaries are sufficiently close to ensure complete filling during the ALD deposition process.

Lastly, we see the biggest potential of the platform as its ability to be integrated with other fabrication techniques previously mentioned. Integrating the platform with twisted  lattices or patterned 2D materials may afford the ability to dynamically open a graphene bandgap~\cite{cao2018unconventional,torres2014introduction}, leading to creation of integrated graphene based emitters in the near infrared~\cite{mikhailov2013graphene}. Although we have integrated this platform with graphene, a myriad of other 2D material systems remain to be explored. The exploration of other 2D materials exhibiting exotic behaviors, such as the in-plane anisotropy of black phosphorus~\cite{cording2024highly}, or with different dominant excitations, such as excitons or plasmons~\cite{basov2016polaritons}, promises both a wealth of interesting physics and new devices, including large-area, integrated, and tunable sources and emitters of exotic electromagnetic responses.

\section*{Acknowledgements}
The authors acknowledge stimulating conversations with Giovanni Scuri, Nicholas Rivera, Marin Soljačić, Ali Ghorashi, Theodore P. Letsou, Charles Roques-Carmes, and Dmitry Kazakov, as well as the support of TEM measurement through Jules Gardener, Adam Graham, and David C. Bell. C.S, S.W.D.L, J.P., P.C, and F.C are supported by NSF Grant under Award Number 2015668. M.R, T.W, and P.K. are supported by Samsung Global Research Program. S.W.D.L. is supported by the Schmidt Science Fellows, in partnership with the Rhodes Trust. S.W.D.L. is also supported by A*STAR Singapore through the National Science Scholarship scheme. I.K. was supported by the Israel Science Foundation (ISF), Grant No. 385/23. M.T. acknowledges the financial support of the European Research Council (ERC) under Grant Agreement No. ERC-2020-STG 948250.

\section*{Author contributions}

MT, CMS, and FC conceived the initial idea. CMS, KV, JSP, and MT developed the fabrication method. MR and CMS fabricated the Hall bar and performed transport measurement. CMS, MR, TW, SWDL, GS, PC and MT performed data analysis and modeling. IK, PK, FC, and MT supervised the project. CMS wrote the paper with inputs from all authors.

\section*{Conflict of interest}
The authors declare no conflict of interest.


\bibliography{references}
\end{document}